\documentclass[12pt]{article}
\usepackage{amssymb,amsmath}
\usepackage{latexsym}
\numberwithin{equation}{section}

\newcommand{\be}{\begin{equation}}
\newcommand{\ee}{\end{equation}}
\newcommand{\bea}{\begin{eqnarray}}
\newcommand{\eea}{\end{eqnarray}}

\def\ti{\widetilde}

\begin{document}

\begin{titlepage}

\begin{flushright}
IPhT-T11/147
\end{flushright}

\bigskip
\bigskip
\centerline{\Large \bf Imaginary Soaring Branes:}
\smallskip
\centerline{\Large \bf A Hidden Feature of Non-Extremal Solutions }
\medskip
 \bigskip
\centerline{{\bf Iosif Bena$^1$, Cl\'{e}ment Ruef$^{\, 2}$ and Nicholas P. Warner$^3$,}}
\bigskip
\centerline{$^1$ Institut de Physique Th\'eorique, }
\centerline{CEA Saclay, CNRS-URA 2306, 91191 Gif sur Yvette, France}
\bigskip
\centerline{$^3$ Department of Physics and Astronomy}
\centerline{University of Southern California} \centerline{Los
Angeles, CA 90089, USA}
\bigskip
\centerline{$^3$ Max Planck Institute for Gravitation, Albert Einstein Institute}
\centerline{Am M\"uhlenberg 1, 14476 Golm, Germany}
\bigskip
\centerline{{\rm iosif.bena@cea.fr,~clement.ruef@aei.mpg.de,~warner@usc.edu } }
\bigskip
 
\bigskip \bigskip

\begin{abstract}
A key property of many BPS solutions of supergravity is the fact that certain probe branes placed in these solutions feel no force, essentially because electric repulsion and gravitational attraction balance one another.  In this letter we show that the existence of brane probes that feel no force is also a property of many non-supersymmetric, non-extremal solutions of supergravity.  This observation requires a new class of brane probes that move with constant velocity along one or several internal directions of the solution but the zero-force condition that makes the branes ``float along'' at constant speed, or soar,  requires the velocity to be purely imaginary.  While these probes are not physical, their no-force condition implies the existence of hidden relations between the warp factors and electric potentials of non-extremal solutions in certain duality frames, and this provides insight into the structure of such solutions and can greatly simplify the search for them.

\end{abstract}

\end{titlepage}
%%%%%%%%%%%%%%%%%%%%%%%%%%%%%%%%%%%%%

% \tableofcontents

%%%%%%%%%%%%%%%%%%%%%%%%%%%%%%%%%%%%%
\section*{Introduction}
\label{sect:Intro}
%%%%%%%%%%%%%%%%%%%%%%%%%%%%%%%%%%%%%

The existence of floating, or mobile, branes is a well-established feature of supersymmetric solutions, and can be traced back to the fact that these solutions saturate a BPS bound.  Specifically, a suitably-chosen probe BPS brane feels an electric repulsion equal to the gravitational attraction, and has a constant Lagrangian regardless of its position. Indeed, one can start from two supersymmetric branes in the region of the moduli space where their gravitational back-reaction can be ignored, and one always finds that, to linear order in the interaction parameter, their electrical and gravitational forces cancel. The fact that this cancellation persists after taking back-reaction into account indicates that the self-interaction of the branes does not increase the mass-to-charge ratio.

Since non-extremal, non-BPS solutions have masses bigger than the BPS bound, it is clear that they will attract any physical brane whether it is BPS, or not\footnote{If the configuration has angular momentum then there may well be a preferred equilibrium position or orbit for the brane.}.   However, one may try to use an unphysical brane-probe action, with a mass that is smaller than the charge, and try to see whether the reduction of the mass of the probe can compensate the increase of the mass of the non-extremal solution. If one starts again with two non-back-reacted systems, one with mass bigger than the charge and one with mass smaller than the charge, the electric and gravitational forces will indeed cancel at first order in the interaction. However, upon taking into account the gravitational back-reaction of the center with mass greater than the charge, this cancellation does not necessarily continue, essentially because the binding energy of the non-extremal object depends non-trivially on the gravitational coupling. Hence, the standard lore is that in a non-extremal solution a probe at a generic location will always feel a force. For example, if one takes  the Reissner-Nordstr\"om  black hole of mass $M$ and charge $Q$ in four dimensions, one finds that the action of a probe of charge $q$ and mass $m$ can never be made constant. Indeed, setting the action to be constant:
\begin{equation}
\nonumber
S = - m \sqrt{-g_{tt}} + q A_t = -m \sqrt{1-{2 M \over r} + {Q^2 \over r^2}} + {q Q \over r} = k 
\end{equation}
implies $m=q$ and $M=Q$, and hence both objects must be BPS.

The purpose of this letter is to show that upon embedding some families of non-extremal solutions in string theory, or in eleven-dimensional supergravity, one can find probes whose action is independent of the position and therefore feel no force. Since these generalized floating branes  generically move along the internal directions at constant velocity, we refer to them as ``soaring branes''. The existence of soaring branes implies non-trivial relations between various functions that appear in these families  of non-extremal solutions and these relations are far from obvious if one simply starts from the supergravity equations of motion.  For floating branes, this approach has already proven its efficiency in solving these equations, not only for BPS and extremal non-supersymmetric solutions  \cite{Goldstein:2008fq,Bena:2009ev,Bena:2009en,Bena:2009qv,Bena:2009fi,Bobev:2009kn,Dall'Agata:2010dy,Galli:2010mg}, but also for flux compactifications \cite{Lust:2008zd,Held:2010az}.  We can thus expect that the more general relations we discover from soaring branes will be of great help for the more complicated non-extremal solutions.

We begin with a very simple example to present the basic procedure:  A soaring brane in the background of an M2-M2-M2 non-extremal black hole in eleven dimensions.   We go on to examine some much more complicated examples and show that some of the known co-homogeneity-two solutions, {\it i.e.} depending on two variables, also admit soaring branes: the JMaRT solution \cite{Jejjala:2005yu}, the charged rotating black ring of Elvang, Emparan and Figueras \cite{Elvang:2004xi}, and the Rasheed-Larsen black hole \cite{rasheed-larsen}.

%%%%%%%%%%%%%%%%%%%%%%%%%%%%%%%%%%%%%
\section{The non-extremal M2-M2-M2 black hole}
\label{sect:M2cubed}
%%%%%%%%%%%%%%%%%%%%%%%%%%%%%%%%%%%%%

We consider the simplest three-charge black hole in eleven-dimensional supergravity compactified on a six-torus.  This solution is given by:
\begin{eqnarray}
 ds_{11}^2 &=& -D^{-2} K dt^2 + D \left( \frac{dr^2}{K} + r^2 d\Omega_3^2 \right) +\sum_{I=1}^3  \frac{D}{D_I} ds_I^2 \,, \label{BHmet} \\
 A^{(3)} &=& \sum_{I=1}^3  \coth\alpha_I \left( 1 - \frac{1}{D_I} \right) \wedge dT_I \,,  \label{BHform}
\end{eqnarray}
where $D \equiv (D_1 D_2 D_3)^{1/3}$,  $ds^2_I$ and $dT_I$ is the flat metric and  volume form  of the $I^{\rm th}$  2-torus and one has:
\begin{equation}
 K \equiv 1 - \frac{r_H^2}{r^2} \,, \qquad \qquad D_I \equiv 1 + \frac{r_H^2}{r^2} \sinh^2\alpha_I \,.
\end{equation}
The mass, charges and temperature of this black hole are 
\begin{eqnarray} 
 G_5 M =  \sum_I M_I  = \frac{r_H^2}{6} \sum_I \cosh 2\alpha_I \,, \quad \quad Q_I = 2\pi^2 r_H^2 \sinh 2\alpha_I \,, \quad \quad  T = \frac{1}{2\pi r_H \prod_I \cosh\alpha_I} \,. \label{Mass}
\end{eqnarray}

We find it useful to rewrite this in a form that will parallel our discussion in Sections 2 and 3 and that is similar to the Ansatz  for  supersymmetric black holes:
\begin{eqnarray}
 ds_{11}^2 &=& -Z^{-2} (dt+k)^2 + Z ds_4^2 + \frac{Z}{Z_I} ds_I^2 \,, \nonumber \\ 
  A^{(3)} &=& \sum_I A_I  \wedge dT_I =  \sum_I  \left( \left( \lambda_I - \frac{1}{W_I} \right) dt + B_I \right) \wedge dT_I \label{metricandC3} \,,
\end{eqnarray}
where $B_I$ has no component along $dt$. The simple black hole defined by (\ref{BHmet}) and  (\ref{BHform})  is not rotating and has no magnetic charges, thus $k=0$ and $B_I=0$. One then has:
\begin{equation}
 Z_I = \frac{D_I}{\sqrt{K}} \,, \quad  Z = (Z_1 Z_2 Z_3)^{1/3}\,, \quad W_I = \frac{D_I}{\coth\alpha_I}\,, \quad \  \lambda_I = \coth \alpha_I \,.
\end{equation}
Unlike the extremal solution, the  base-space metric
\begin{eqnarray}
  \quad ds_4^2 = \frac{dr^2}{\sqrt{K}} + \sqrt{K} d\Omega_3^2 
\end{eqnarray}
is not flat and  $Z_I$ and $W_I$ are not equal. However,  they are related:
\begin{equation}
 \frac{1}{Z_I^2} = \frac{1}{W_I^2} - \frac{2}{\sinh 2\alpha_I} \frac{1}{W_I} \,. \label{hidden}
\end{equation}
As we will see below, this relation between the warp factors $Z_I$ and the electric potentials $W_I$ is essentially responsible for the existence of soaring branes. 

The action for a probe M2-brane is:
\begin{eqnarray}
 S = \int d^3z \sqrt{-\det{g_{\rm induced}}} + \int A^{(3)} \,.
\end{eqnarray}
A generic, physical brane will be attracted by the black hole. If the brane is  wrapped along   first torus and is boosted with velocity $v$ along the second torus, its action becomes:
\begin{equation} \label{action1}
 S = \int d^3z \sqrt{\frac{D^2}{D_1^2}\left( \frac{K}{D^2} - v^2 \frac{D}{D_2} \right)} + \coth \alpha_1 \left( 1 - \frac{1}{D_1} \right) \,.
\end{equation}
As one would expect, this action is, in general, a non-trivial function of $r$ and does not have a physical minimum:  Such a physical probe brane will thus fall into the black hole.  However, a short manipulation shows that if $v$ takes the  {\it imaginary} value
\begin{equation}
v= v_\pm \equiv \frac{i}{ \sinh (\alpha_1 \mp \alpha_3 )}  \,
\label{vvalue}
\end{equation}
then the action is constant,
\begin{equation}
S = u_\pm =  \coth  (\alpha_1 \mp \alpha_3 )  \,,
\label{uvalue}
\end{equation}
and the brane floats along, or soars! It is interesting to observe  that the velocity of the soaring brane wrapped on the first torus and moving on the second one does not depend on the second torus-parameter $\alpha_2$, but depends on the parameter of the third torus $\alpha_3$. We also note that expression for the velocity is regular when $\alpha_3 \to \mp \alpha_1$. 

This probe brane only becomes physical when its velocity to goes to zero and from \eqref{vvalue} one sees that this only happens for $\alpha_I \to \infty$. For the background fields to stay finite in this limit one must also  take $r_H \to 0$, keeping $ r_H \cosh \alpha_I $ fixed.   To get a black hole with finite horizon area one must take this limit for all three $\alpha_I $ and then, from \eqref{Mass}, one sees that this is exactly the limit where the black hole becomes extremal. One thus recovers the standard floating probe BPS M2 brane in the background of the BPS black hole.  

Hence, the non-extremal M2-M2-M2 black hole admits soaring M2-branes. Clearly this brane, with its imaginary velocity,  is not a physical object, but the  constancy of its action implies the existence of a hidden relation between the parameters of the black hole solution. For $\alpha_3 \to \pm \alpha_1$, it can be precisely traced back to (\ref{hidden}). 

As anticipated, because the black hole has mass bigger then charge, the brane that feels no force should have the mass smaller than the charge, and this comes because the boost velocity along the torus is imaginary and the $\gamma$-factor $\sqrt{1-v^2} = u$ is bigger than 1. This can be made more precise: the boost increases of the charge-to-mass ratio of the probe to match precisely the mass to charge ratio of the black hole. If $M_I$ denotes  the contribution to the black-hole mass from the $I^{\rm th}$ brane (\ref{Mass}), then (taking the soaring velocity to be $v_+$):
\begin{eqnarray} 
 \sqrt{1-v^2} (Q_1 + Q_3) &=& 2\pi^2 r_H^2 \coth  (\alpha_1 + \alpha_3 ) \,  \big(  \sinh 2\alpha_1 + \sinh 2\alpha_3 \big)  \nonumber \\
 &=& 2\pi^2 r_H^2 (\cosh 2\alpha_1 + \cosh 2\alpha_3)  ~=~ 12\pi^2 G_5 (M_1 + M_3)  \,. \label{MQcompensation}
\end{eqnarray}

We now consider some much more sophisticated examples of soaring branes that demonstrate that the result here is not merely an accident of simplicity or spherical symmetry, but an illustration of a much richer and more complex phenomenon.

%%%%%%%%%%%%%%%%%%%%%%%%%%%%%%%%%%%%%
\section{The JMaRT solution}
\label{sect:JMaRT}
%%%%%%%%%%%%%%%%%%%%%%%%%%%%%%%%%%%%%

The JMaRT solution  \cite{Jejjala:2005yu} is a highly non-trivial bubbled, non-extremal two-centered solution that also has soaring branes.    Using \cite{Gimon:2007ps}, one can write the  metric in the standard base-fiber decomposition 
 (\ref{metricandC3}) with warp factors and vector potentials\footnote{We will not need the expressions of the rest of the fields, and thus just refer the reader to \cite{Jejjala:2005yu,Gimon:2007ps} for the complete details.}
\bea
 Z_I &=& \frac{\tilde H_I}{2c \sqrt{V}}  \,, \label{ZIjmart}\\ 
 A_I &=& -W_I^{-1} dt ~+~ B_I ~=~ \tilde H_I^{-1} Q_I \, dt ~+~ B_I \,,  \label{Aiform1} \\
  {\rm{with}} & & \ V \equiv \frac{f^2}{c^2} -p \,, \qquad {\tilde H_I} = \frac{f + c E_I}{2}\,,
\eea
where $c$ is a constant, $B_I$ is a one-form on the four-dimensional spatial base, and we have introduced the function $f$ and the parameter $p$:
\begin{equation}
f  ~\equiv~  (r+r_c)  ~+~  (m^2-n^2) (r-r_c) \,, \qquad p  ~\equiv~  (m^2-n^2)^2   ~-~  2\,(m^2 + n^2)  + 1 \,,
\label{pfdefn}
\end{equation}
with
\begin{equation}
r_c^2  \equiv r^2 + 2 r c \cos \theta + c^2\,.
\label{rcdefn}
\end{equation}
 The parameters $E_I$ in the warp factors are related to the charges via:
\begin{equation}
E_I \equiv 
\sqrt{ \frac{Q_I^2}{4\, c^2}   ~+~  p  } \,,
\end{equation}
and this insures that the functions $\tilde H_I$ are everywhere positive \cite{Jejjala:2005yu}. The parameter $p$ controls the supersymmetry breaking: if $p=0$ then $E_I = \frac{Q_I}{2c}$  and the solution is BPS, as evident from (\ref{MADM}).

The absence of closed time-like curves requires that $m$  and $n$ must both be integers.  
The ADM mass and angular momenta of the five-dimensional asymptotically flat solution  
(in units where $4G^{(5)}/\pi=1$) are given by:
\begin{equation}
M_{ADM} ~=~ 2\,c \,(E_1 + E_2 + E_3), 
\label{MADM}
\end{equation}
and
\begin{equation}
J_\varphi  ~=~ \frac{1}{\sqrt{m\, n}}\,   (n-m) \,  \sqrt{Q_1 Q_2 Q_3}\,, \qquad J_\tau  ~=~ \frac{1}{\sqrt{m\, n}}\,   (n+m) \,  \sqrt{Q_1 Q_2 Q_3} \,.
\end{equation}

This non-extremal solution depends on two variables, and because of (\ref{rcdefn}) it can be thought of as a two-center solution. The existence of branes that feel no force is therefore a much more unexpected feature. However, once again, $Z_I$ and $W_I$ (defined in \eqref{ZIjmart}-\eqref{Aiform1}) are related:
\bea \label{ZWjmart}
 \frac{1}{Z_I^2} = \frac{1}{W_I^2} + \frac{E_I}{c \,Q_I W_I} + 1 \,.
\eea
An M2 brane wrapping the first torus and moving at constant speed $v$ around the second one\footnote{Again, these choices for the tori are completely arbitrary, due to the permutation symmetry of the solution.}. The action for this brane is 
\begin{equation}
S   ~=~  \sqrt{\frac{\tilde H^2}{\tilde H_1^2} \bigg(\frac{1}{Z^2} ~-~ v^2 \, \frac{\widehat H}{\tilde H_2}   \bigg)}  ~-~  \frac{Q_1}{\tilde H_1}  \,.
\label{DBI1}
\end{equation}
where $\tilde H \equiv  (\tilde H_1 \tilde H_2 \tilde H_3)^{1/3}$.
For the brane to soar, this action must be independent of the two variables on which the solution depends. This happens when 
\begin{equation}
v  ~=~ \sin \alpha \,, 
\label{alphadefn}
\end{equation}
where
\begin{equation}
\cos  \alpha ~\equiv~ - \frac{(Q_1 \pm Q_3)}{2\, c\, (E_1 -E_3)} ~=~-  \frac{2\, c\, (E_1 + E_3) }{(Q_1 \mp Q_3)}   \,,
\label{quadsol1}
\end{equation}
The value of the action is then simply: $S = \cos \alpha$. Since $Q_\ell >0$ and $2 c E_ \ell \ge Q_\ell >0$, with equality if and only if $p=0$, it is clear that the magnitude of the right-hand side of (\ref{quadsol1}) is always greater than or equal to $1$, and thus $v$ is imaginary.  As a result, one can only find a real solution if $p=\alpha=0$. This is the supersymmetric floating brane.

On the other hand, just like the non-extremal black hole, when the velocity, $v$, is imaginary, we can have soaring branes, whose increase in charge with respect to the BPS bound compensates for the fact that the background has a charge smaller than its mass.  Once again, it is obvious from \eqref{alphadefn} and \eqref{quadsol1} that the $\gamma$-factor associated to $v$ is exactly the one needed to balance the mass-to-charge ratio:
\begin{eqnarray}
 \sqrt{1-v^2}\, (Q_1 + Q_3) = M_1 + M_3 \,. \label{JMaRTsoaring}
\end{eqnarray}
%

%%%%%%%%%%%%%%%%%%%%%%%%%%%%%%%%%%%%%
\section{The non-BPS black ring}
\label{sect:nBPSBR}
%%%%%%%%%%%%%%%%%%%%%%%%%%%%%%%%%%%%%

There are also soaring branes in several sub-families of the non-BPS black rings  found in \cite{Elvang:2004xi}. To demonstrate this  we first write these solutions in the canonical form  (\ref{metricandC3}) with warp factors, base metric and vector potentials:
\begin{eqnarray}
 Z_I &=& \frac{h_I}{\sqrt{U_I}} \,, \\ 
 A_I &=&  (1 -W_I^{-1} ) \, dt ~+~ B_I = \frac{(U_I -1)}{h_I} \,  c_I \, s_I \, dt ~+~ B_I \,,  \label{Aiform2}  \\
 ds_4^2 &=& \sqrt{F(x)F(y)H(x)^3 H(y)^3}\frac{R^2}{(x-y)^2} \\ \nonumber
 && \qquad \qquad \qquad \qquad \left[ -\frac{G(y)}{F(y)H(y)^3} d\psi^2 - \frac{dy^2}{G(y)} + \frac{G(x)}{F(x)H(x)^3} d\phi^2 + \frac{dx^2}{G(x)}  \right] \,,
\end{eqnarray}
where $B_I$ is a one-form on the spatial base. The functions appearing in the solution are:
\begin{equation}
  \begin{array}{lclclcl}
  F(\xi) &=& 1 + \lambda\, \xi \, ,&~~~~~~~&
  G(\xi) &=& (1-\xi^2)\,(1+\nu\, \xi)\, , ~~~~~~\\[2mm]
  H_I(\xi) &=& 1- \mu_I\, \xi \, ,&~~~~~~&
  H(\xi) &\equiv& \big[ H_1(\xi) H_2(\xi) H_3(\xi) \big]^{1/3}\, ,
  \end{array}
\label{FGHdefns}
\end{equation}
and it is convenient to define the functions:
\begin{equation}
  U_I~\equiv~   \frac{F(y) H(x)^3}{F(x) H(y)^3}
 \frac{H_I(y)^2}{H_I(x)^2} \,, \quad h_I ~\equiv~  {{c_I}}^2 - U_I  {{s_I}}^2\,, \quad W_I = \frac{h_I}{(c_I+s_I)(c_I - U_I s_I)} \,,
\label{UI}
\end{equation}
where
\begin{equation}
c_I\equiv \cosh\alpha_I\,,\qquad s_I\equiv \sinh\alpha_I\,.
\end{equation}
As in the earlier examples, there is also a hidden relation, analogous to \eqref{hidden} and \eqref{ZWjmart},   between the metric warp factors and the electric potentials:
\be \label{ZWblackring}
 \frac{1}{Z_I^2} = \frac{1}{W_I^2} + \frac{(c_I + s_I)^2}{c_I s_I}\left( -\frac{1}{W_I} + 1 \right) \,.
\ee
One can repeat the calculations of the previous solution to find soaring branes but, unlike the other examples, this only happens for some specific choices of parameters. We still take the M2-brane to be wrapped on the first torus and boosted along the second with velocity, $v$.  The probe action is then:
\be
 S = \int d^3z \sqrt{\frac{Z^2}{Z_1^2}\left( \frac{1}{Z^2} - v^2 \frac{Z}{Z_2} \right)} + \frac{U_1 - 1}{h_1} c_1 s_1\,.
\end{equation}
Setting that the action is equal to a constant, $u$,  one finds the condition:
\be \label{constraintU1U3}
 U_1 - v^2 \sqrt{\frac{U_1}{U_3}} (c_1^2 - U_1 s_1^2)(c_3^2 - U_3 s_3^2) - (U_1 s_1(c_1 - u s_1) + c_1(c_1 u - s_1) )^2 = 0 \,.
\end{equation}
For generic parameters, $U_1$ and $U_3$ are distinct functions and the constraint \eqref{constraintU1U3} cannot be solved, and thus  there are no soaring branes.  However, if one imposes certain restrictions on the parameters to relate $U_1$ and $U_3$, floating branes can still exist. The first (and obvious) restriction is $\mu_1=\mu_3$, such that $U_1=U_3$.   But there exists also a second (less obvious) one: if one takes $\lambda = - \mu_2$, then from \eqref{UI} one obtains that $U_3 = U_1^{-1}$.

Supppose one sets $\mu_1 = \mu_3$ then one has $U_1=U_3$, and then \eqref{constraintU1U3} can be solved by 
\begin{eqnarray}
 u_\pm = \coth(\alpha_1 \pm \alpha_3)  \,, \qquad v_\pm =  \frac{i}{\sinh(\alpha_1 \pm \alpha_3)} \,.
\end{eqnarray}
%
%There is again a free sign, that we take to be $+$.
This is somewhat reminiscent of the soaring branes for the three-charge black hole for which one has \eqref{vvalue} and \eqref{uvalue}.  The soaring brane has again imaginary velocity, and exists for any value of the extremality parameter, $\nu$.  The first and third charges of the solution are 
\begin{eqnarray}
 Q_1 = \sinh 2\alpha_1\,  Q \,, \quad Q_3 = \sinh 2\alpha_3\, Q \,, \quad  {\rm{with}} \ Q \equiv \frac{R^2}{1-\nu} ( \lambda + \mu_2 ) (1+ \mu_1)^2 \,,
\end{eqnarray}
and one can again check that
\begin{eqnarray}
 \sqrt{1-v^2}(Q_1 + Q_3) &=& \frac{\cosh(\alpha_1 + \alpha_3)}{\sinh(\alpha_1 + \alpha_3)} ( \sinh 2\alpha_1 + \sinh 2\alpha_3 ) \, Q \\ \nonumber
 &=& ( \coth 2\alpha_1 + \coth 2\alpha_3 ) Q = M_1 + M_3 \,.
\end{eqnarray}

The other solution has  $\lambda = -\mu_2$, which implies  $U_3=U_1^{-1}$, and then equation \eqref{constraintU1U3} can be solved by 
\begin{eqnarray}
 u_\pm = \frac{\sinh(\alpha_1 \pm \alpha_3)}{\cosh(\alpha_1 \pm \alpha_3)} \,, \qquad v_\pm = \frac{1}{\cosh(\alpha_1 \pm \alpha_3)} \,.
\end{eqnarray}
There is again a free sign, which we take to be $+$. The velocity appears, at first glance, to be real, which would contradict our intuition that a regular non-extremal solution should only admit imaginary soaring branes. However the black-ring solution is only regular for 
\begin{eqnarray}
 0 < \nu \leq \lambda < 1 \,, \quad 0 \leq \mu_I < 1 \,,
\end{eqnarray}
and hence the parameter choice for which  soaring branes exist, $\mu_2=-\lambda$, does not lie within the physical range of solutions. 

Finally, one can look at the relation between the mass and the charges of the solution. For $\mu_2=-\lambda$, the first and third charges are given by
\begin{eqnarray}
 Q_1 = \sinh 2\alpha_1 \,Q \,, \quad Q_3 = - \sinh 2\alpha_3 \, Q \,, \quad {\rm{with}} \ Q \equiv \frac{R^2}{1-\nu} (1 - \lambda)^2(\mu_3 - \mu_1) \,.
\end{eqnarray}
Using these expressions, one can again check that
\begin{eqnarray}
 \sqrt{1-v^2}(Q_1 + Q_3) &=& \frac{\sinh(\alpha_1 + \alpha_3)}{\cosh(\alpha_1 + \alpha_3)} (\sinh 2\alpha_1 - \sinh 2\alpha_3) Q \nonumber \\ 
 &=& (\coth 2\alpha_1 - \coth 2\alpha_3) Q = M_1 + M_3 \,.
 \label{BRsoaring}
\end{eqnarray}
Hence, the velocity of the probe brane is real, and its mass is bigger than its charge. However, as we have noted above, the black ring background is not physical, and has mass smaller than charge. Thus, the soaring brane exists again because an increase of the mass is compensated by the increase of a charge, but now it is the background whose charge is bigger than the BPS bound.

%%%%%%%%%%%%%%%%%%%%%%%%%%%%%%%%%%%%%
\section{The Rasheed-Larsen black hole}
\label{sect:RLBH}
%%%%%%%%%%%%%%%%%%%%%%%%%%%%%%%%%%%%%

The solutions presented in the previous section were all written in an M-theory frame where the  black hole electric charges correspond to three sets of M2 branes and the soaring brane is an M2 brane wrapped on one torus and moving along another one with an imaginary velocity.  We now focus on a rather different kind of solution with a different kind of soaring brane.
The general Rasheed-Larsen black hole is a dyonic D0-D6 solution \cite{rasheed-larsen}, whose under-rotating limit 
\cite{Astefanesei:2006dd, Emparan:2007en} can be dualized \cite{Bena:2009ev} to an ``almost-BPS'' rotating black hole, which admits floating M2-branes. It is natural to ask whether the non-extremal version of the solution admits a type of soaring brane, and whether this brane is physical or has imaginary parameters. 

%%%%%%%%%%%%%%%%%%%%%%%%%%%%%%%%%%%%
\subsection{D0-D6 solution in the IIA framework}
%%%%%%%%%%%%%%%%%%%%%%%%%%%%%%%%%%%%

We write the Rasheed-Larsen solution in a type IIA frame by starting from the form in five dimensions given in \cite{rasheed-larsen} then adding a trivial six-torus, and then reducing along the fiber of the solution. In the IIA frame, the solution is given by:
\begin{eqnarray}
 ds^2_{10} &=& -\frac{\Delta_\theta}{\sqrt{H_p H_q}} \, (dt + {\bf B})^2 + \frac{\sqrt{H_p H_q}}{\Delta_\theta}\left( \frac{\Delta_\theta}{\Delta}dr^2 + \Delta_\theta d\theta^2 + \Delta d\phi^2 \right) + \sqrt{\frac{H_q}{H_p}} \, ds_{T^6}^2 \,, \nonumber \\
 \rm{e}^{2\Phi} &=& \left( \frac{H_q}{H_p} \right)^{3/2} \,,
\end{eqnarray}
for the NS-NS fields and 
\begin{eqnarray}
 C^{(1)} = {\bf A} \,, \quad  C^{(7)} = \ti{\bf A} \wedge dT_1 \wedge dT_2 \wedge dT_3 \,,
\end{eqnarray}
for the R-R fields. The functions appearing in the solution are
\begin{eqnarray}
 H_p &=& r^2 + \alpha^2 \cos^2\theta + r(p-2m) + \frac{p}{p+q}\frac{(p-2m)(q-2m)}{2}  \\ \nonumber
 && \qquad \qquad \qquad + \frac{p}{2(p+q)}\sqrt{(p^2-4m^2)(q^2-4m^2)} \frac{\alpha}{m} \cos\theta \,, \\ 
 H_q &=& r^2 + \alpha^2 \cos^2\theta + r(q-2m) + \frac{q}{p+q}\frac{(p-2m)(q-2m)}{2}  \\ \nonumber
 && \qquad \qquad \qquad - \frac{q}{2(p+q)}\sqrt{(p^2-4m^2)(q^2-4m^2)} \frac{\alpha}{m} \cos\theta \,, \\
 \Delta_\theta &=& r^2 + \alpha^2 \cos^2\theta - 2 m r \,, \qquad 
 \Delta ~=~ r^2 + \alpha^2 - 2mr \,,
\end{eqnarray}
and the one-forms defining the angular momentum and the R-R fields are
\begin{eqnarray}
 {\bf B} &=& \sqrt{pq}\frac{(pq + 4 m^2)r - m(p-2m)(q-2m)}{2(p+q)\Delta_\theta} \frac{\alpha}{m} \sin^2\theta \, d\phi \,, \\ 
 {\bf A} &=& -\frac{1}{H_q}\left( 2 Q \big( r + \frac{p-2m}{2} \big) - \sqrt{\frac{q^3(p^2-4m^2)}{4(p+q)}} \frac{\alpha}{m} \cos\theta \right) \, dt - \frac{1}{H_q} \Big( 2P (H_q + \alpha^2\sin^2\theta)\cos\theta  \nonumber \\
 && - \sqrt{\frac{p(q^2-4m^2)}{4(p+q)^3}}((p+q)(p r - m(p-2m))+q(p^2-4m^2))\frac{\alpha}{m}\sin^2\theta \Big) \, d\phi \,, \\ \nonumber
 \ti{\bf A} &=& \frac{1}{H_p}\left( 2 P \big( r + \frac{q-2m}{2} \big) + \sqrt{\frac{p^3(q^2-4m^2)}{4(p+q)}} \frac{\alpha}{m} \cos\theta \right) \, dt - \frac{1}{H_p} \Big( 2Q (H_p + \alpha^2\sin^2\theta)\cos\theta  \nonumber \\
 && + \sqrt{\frac{q(p^2-4m^2)}{4(p+q)^3}}((p+q)(q r - m(q-2m))+p(q^2-4m^2))\frac{\alpha}{m}\sin^2\theta \Big) \, d\phi \,. 
\end{eqnarray}
The parameters $p$, $q$, $\alpha$, and $m$ are restricted to the range $p \geq 2m$, $q \geq 2m$, $m \geq |\alpha|$ for the solution to be physical, and they are related to the four-dimensional physical parameters of the solutions $M$, $J$, $P$ and $Q$ by
\begin{eqnarray}
 G_4 M \!\!&=&\!\! \frac{p+q}{2} \,, \qquad  G_4 J = \sqrt{pq}\frac{pq + 4 m^2}{4(p+q)}\frac{\alpha}{m} \,, \\ \nonumber
 Q  \!\!&=&\!\! \sqrt{\frac{q(q^2-4m^2)}{4(p+q)}} \,, \qquad  P = \sqrt{\frac{p(p^2-4m^2)}{4(p+q)}} \,.
\end{eqnarray}
The $C^{(7)}$ RR-potential has been computed from the equation of motion of $C^{(1)}$, and it is quite natural that the result is given by the same expression as $C^{(1)}$ with $ p \leftrightarrow q$ and $ \alpha \leftrightarrow - \alpha $. 

There are two extremal limit to this solution, the under-rotating one, $m \to 0$, $\alpha \to 0$ keeping $\alpha / m $ fixed, and the over-rotating one, taking $\alpha=m$.

%%%%%%%%%%%%%%%%%%%%%%%%%
\subsection{Soaring D6-brane}
%%%%%%%%%%%%%%%%%%%%%%%%%

We are looking for a floating brane in the background of the Rasheed-Larsen solution. In the previous sections, our branes were boosted M2 branes along some directions of the internal six-torus. Using the duality between almost-BPS solutions and the under-rotating Rasheed-Larsen solution \cite{Bena:2009ev}, we expect that the soaring brane should be a D6-brane stretched along the six internal directions, with abelian world-volume fluxes that give rise to D4, D2, D0 and F1-charges. We choose for simplicity, a probe, D6 brane with world-volume fluxes $\mathcal F_{12}=\mathcal F_{34}=\mathcal F_{56}=F$, and F1-charge along $x_2$ obtained by turning on $\mathcal F_{tx_2}=G$. 

The D6-brane action is
\begin{eqnarray}
 S \!\!&=&\!\! S_{DBI} + S_{WZ} \\ \nonumber 
 &=&\!\! \int \!\! d^7z \, {\rm{e}}^{-\Phi} \sqrt{-\det(g+\mathcal F)} ~+~   \int \!\! \big[\, C^{(7)}_t + F^3 \, C^{(1)}_t \, \wedge\, \big(\wedge_{I=1}^6 dx_I \big)\, \big] \,. 
\end{eqnarray}
%
%With our choices%, the matrix $g+\mathcal F$ is given by
%
%\begin{eqnarray}
% g+\mathcal F = \left[ \begin{array}{ccccccc}
% -\frac{\Delta_\theta}{\sqrt{H_p H_q}} & 0 & G & 0 & 0 & 0 & 0 \\
% 0 & \sqrt{\frac{H_q}{H_p}} & F & 0 & 0 & 0 & 0 \\
% -G & -F & \sqrt{\frac{H_q}{H_p}} & 0 & 0 & 0 & 0 \\
% 0 & 0 & 0 & \sqrt{\frac{H_q}{H_p}} & F & 0 & 0 \\
% 0 & 0 & 0 & -F & \sqrt{\frac{H_q}{H_p}} & 0 & 0 \\
% 0 & 0 & 0 & 0 & 0 & \sqrt{\frac{H_q}{H_p}} & F \\
% 0 & 0 & 0 & 0 & 0 & -F & \sqrt{\frac{H_q}{H_p}}  
% \end{array}
% \right]
%\end{eqnarray}
%
With our choices, it becomes
\begin{eqnarray}
 S_{DBI} \!\!&=&\!\! \int \!\! d^7z \, \frac{H_q + F^2 H_p}{H_p H_q} \sqrt{\Delta_\theta ( H_q + F^2 H_p) - G^2 H_p H_q } \,,  \nonumber \\
 S_{WZ} \!\!&=&\!\! \int \!\! d^7z \, \frac{1}{H_q H_p}\left( 2 H_q P \big( r + \frac{q-2m}{2} \big) - 2 F^3 \, H_p Q \big( r + \frac{p-2m}{2} \big) \right. \\ \nonumber
 && \!\! \left. + \left( H_q \sqrt{\frac{p^3(q^2-4m^2)}{4(p+q)}} + F^3 \, H_p \sqrt{\frac{q^3(p^2-4m^2)}{4(p+q)}} \right) \frac{\alpha}{m} \cos\theta \right) \,.
\end{eqnarray}
Given than the expression above is a complicated function of $r$ and $\theta$, it seems rather unlikely that there exists a choice of fluxes for which this becomes a constant. However, a careful investigation shows that for 
\begin{eqnarray} \label{FGvalues}
 F^2 = \frac{p(q^2 - 4m^2)}{q(p^2-4m^2)} \,, \qquad G^2 = -\frac{4(p+q) m^2}{q(p^2-4m^2)} \,,
\end{eqnarray}
the expression under the square root in the DBI action becomes a perfect square, and 
\be
S_{DBI} = \int \!\!\! d^7z  \sqrt{\frac{p}{(p+q)(p^2-4m^2)}} \frac{H_q + F^2 H_p}{H_p H_q} \left( (p+q)(r^2 + \alpha^2 \cos^2\theta ) + (p-2m)(q-2m)(r-m)\right)
\end{equation}
giving a constant total action
\be
 S_{DBI} + S_{WZ} =  \frac{\sqrt{p}(p+q)^{3/2}(p q - 4 m^2)}{q(p^2-4m^2)^{3/2}} \,.
\end{equation}
We have therefore found a floating brane in the non-extremal Rasheed-Larsen solution, for all values of the parameters of this solution. For the parameter range in which this solution describes a physical black hole ($p \geq 2m$) the electric field on the probe brane $G$ \eqref{FGvalues} has to be purely imaginary, and thus the floating brane has mass smaller than charge. Since the string flux is dual to a boost charge in the M2-brane duality frame, this parallels the story in the previous sections. 

The imaginary electric field of the soaring brane vanishes when $m \to 0$ or when $p \to \infty$. In the second limit the physical parameters of the black hole also diverge, and so the resulting solution is not meaningful. In the first limit, for physical quantities to stay finite we also have to take $\alpha \to 0$, keeping $\alpha/m=J$ fixed. This is exactly the under-rotating extremal limit, in which the solution is dual to the almost-BPS solution constructed in \cite{Bena:2009ev}, and the floating, fluxed D6 of this solution becomes the floating M2 of the almost-BPS solution.

%%%%%%%%%%%%%%%%%%%%%%%
\section{Discussion}
\label{sect:Conc}
%%%%%%%%%%%%%%%%%%%%%%%

We have discovered that there exist probe branes that have a constant (imaginary) speed or electric flux, and whose action is position-independent when placed in certain highly-complicated solutions of four- and five-dimensional supergravity embedded in string theory. These imaginary probes are unphysical in that they have mass smaller than charge, and at linear order in the gravitational interaction one may indeed expect that a non-extemal brane or black hole with a mass larger than the charge could exert no force on such a probe. However, for a fully-back-reacted solution, and especially for rotating solutions that depend on two variables, the fact that such soaring branes exist is highly unexpected. 

This indicates that there exist simple algebraic relations between the functions that appear in non-extremal solutions. Such relations exist usually in supersymmetric solutions (where they follow from the Killing spinor variation) or in extremal solutions  \cite{Goldstein:2008fq,Bena:2009ev,Bena:2009en,Bena:2009qv,Bena:2009fi,Bobev:2009kn,Dall'Agata:2010dy,Galli:2010mg}, where the supersymmetry is very softly broken\footnote{Note that with such  soft-breaking one can still match microscopic and macroscopic entropies  \cite{Emparan:2006it,Emparan:2007en}.} and where one expects extremality to manifest itself by a factorization of Einstein's equations.  However, the existence of such relations for non-extremal solutions, and the fact that on these solutions Einstein's equations also factorize, is much more unexpected.

For non-extremal single-center solutions that depend on one variable it has been known for some time that there exists a fake superpotential from which one can derive the full solution \cite{Miller:2006ay,Perz:2008kh, Chemissany:2009hq, Chemissany:2010zp}. This also seems  surprising, but upon a careful investigation of the equations satisfied by the functions of one variable that enter the Ansatz, one can argue that the factorization of the Einstein's equations implied by the fake superpotential can indeed be derived from the equations themselves. However, for solutions that depend on two variables fake superpotentials  have never been constructed, and the fact that Einstein's equations can be factorized is much more unexpected.

We can ask whether one can make this non-extremal factorization more precise. One common property of the solutions that admit soaring branes is that the five-dimensional dilaton, or the volume of the six-torus in the eleven-dimensional solution, is constant. One can ask whether one can still find soaring branes in solution with a non-zero dilaton, or, from a lower-dimensional supergravity perspective, with non-trivial hypermultiplet fields turned on. The naive answer is that this is unlikely, because the extra dilaton or the extra hypermultiplet fields enter non-trivially in the probe action and would, in general, spoil the delicate balance needed for the action to become a constant. However, the Rasheed-Larsen solution as well as the solutions obtained by dualizing the black hole, ring and JMaRT solutions to other duality frames have a non-trivial dilaton profile and also admit soaring branes, and hence the non-trivial cancelations needed for a soaring brane may survive the introduction of a dilaton.

The other very important feature of the solutions we investigate is that the warp factors $Z_I$ are related to the electric potentials $W_I$.  In all but the Rasheed-Larsen solution, this relationship, (\ref{hidden}), (\ref{ZWjmart}), (\ref{ZWblackring}), takes the form: 
\be
\frac{1}{Z_I^2} = \frac{1}{W_I^2} + \frac{a_I}{W_I} + b_I \,, \label{BIgeneral}
\end{equation}
which one can rewrite as
\be
 \frac{1}{Z_I^2} + v_I^2 = \left( \frac{1}{W_I} + u_I \right)^2 \,. \label{BIsquare}
\end{equation}
If one dualizes a simplified version of the Rasheed-Larsen solution to the duality frame where it has M2 charges, one finds  exactly the same relation, which leads us to believe that this relation is generic in the appropriate duality frame.

The form of (\ref{BIsquare}) is very suggestive of the cancellation between the Born-Infeld action and the Wess-Zumino action that produces soaring branes. This is indeed true in the simple examples in Sections 1 and 2. However, for the black ring, the relation (\ref{BIgeneral}) is {\it always} true, but to get a soaring brane we needed to set two dipole charges equal and the cancellation was not as straightforward as (\ref{BIsquare}) would imply.  The solutions in  Sections 1 and 2 either have no dipole moments or have all three dipole moments equal, which suggests that  for solutions with arbitrary values of the dipole moments one might need to construct more general soaring branes.
 
While imaginary soaring branes might seem esoteric, they have important physical implications. First, their existence will have a  valuable practical applications when it comes to finding solutions.  As was evident from the simpler, floating-brane Ansatz \cite{Bena:2009fi}, knowing  relationships between warp factors and electrostatic potentials can greatly simplify the search for interesting physical solutions in supergravity and string theory.  The remarkable thing about  imaginary soaring branes is that by using such unphysical probes we can find such relationships in non-extremal solutions that are physically very important. 

In addition, the existence of soaring branes shows that the balancing of gravitational attraction and electrostatic repulsion persists beyond linear order for non-BPS objects. In particular, our results indicate that we have very precise and independent control of the amount of non-extremality coming from each of the charge systems. The non-extremality is determined by the soaring speeds of the branes in relations like (\ref{MQcompensation}), (\ref{JMaRTsoaring}) and (\ref{BRsoaring}). This is also manifest in (\ref{BIgeneral}), which illustrates how each electrostatic potential contributes independently to the geometry. We believe this observation will have interesting consequences for our understanding of the structure of, and the interactions between, non-BPS branes.

%%%%%%%%%%%%%%%%%%%%%%%%%%%%%%%%%%%%%

\bigskip
\section*{Acknowledgments}

\noindent We would like to thank Stefano Giusto and Ilarion Melnikov for useful discussions.  The work of IB was supported in part by the  ANR grant 08-JCJC-0001-0, and by the ERC Starting Independent Researcher Grant 240210 - String-QCD-BH. The work of NPW was supported in part by DOE grant DE-FG03-84ER-40168. NPW is also very grateful to the IPhT, CEA-Saclay for hospitality while much of this work was done. 

\smallskip
%

%%%%%%%%%%%%%%%%%%%%%%%%%%%%%%%%%%%%%

%%%%%%%%%%%%%%%%%%%%%%%%%%%%%%%%%%%%%

%%%%%%%%%%%%%%%%%%%%%
%%%%%%%%%%%%%%%%%%%%
\end{document}